\documentclass[letter]{aa}  

\newcommand{\adleo}{AD\,Leo\,}

\usepackage{graphicx}
\usepackage{txfonts}
\usepackage{bm}
\begin{document} 

   \title{The Great Flare of 2021 November 19 on  AD\,Leo}

  \subtitle{Simultaneous {\em XMM-Newton} and {\em TESS} observations}

   \author{B. Stelzer
          \inst{1,2}
          \and M. Caramazza \inst{1} 
          \and S. Raetz \inst{1} 
          \and C. Argiroffi \inst{2,3} \and M. Coffaro \inst{1}
          }

   \institute{$^1$ Institut f\"ur Astronomie \& Astrophysik, Eberhard-Karls Universit\"at T\"ubingen, 
   Sand 1, 72076 Germany \\ 
   $^2$ INAF - Osservatorio Astronomico di Palermo,  Piazza del Parlamento 1,
  I-90134 Palermo, Italy \\
   $^3$ Department of Physics and Chemistry, University of Palermo,  Piazza del Parlamento 1,
  I-90134 Palermo, Italy \\
              \email{stelzer@astro.uni-tuebingen.de}
   }
         

   \date{Received 30-July-2022; accepted 08-September-2022}

 
  \abstract{We present a detailed analysis of a superflare  on the active M dwarf star AD\,Leonis. The event presents a rare case of a stellar flare observed simultaneously in X-rays (with {\em XMM-Newton}) and in optical (with the {\em Transiting Exoplanet Survey Satellite}, TESS). 
  The radiated energy both in the $0.2-12$\,keV X-ray band 
($1.26 \pm 0.01 \cdot 10^{33}$\,erg)
 and the bolometric value ($E_{\rm F,bol} = 5.57 \pm 0.03 \cdot 10^{33}$\,erg) put this event at the lower end of the  superflare class. The exceptional photon statistics deriving from the proximity of AD\,Leo has enabled  measurements in the $1-8$\,\AA~GOES band  for the peak flux (X1445 class)  
  and integrated energy 
  ($E_{\rm F,GOES} = 4.30 \pm 0.05 \cdot 10^{32}$\,erg), 
  making possible a direct comparison with data on flares from our Sun. 
  From extrapolations of empirical relations for solar flares we estimate that 
  a proton flux of at least $10^5\,{\rm cm^{-2} s^{-1} sr^{-1}}$ 
 accompanied the radiative output. With a
 time lag of $300$\,s between the peak of the TESS white-light flare and the GOES band flare peak as well as a clear Neupert effect 
 this event follows very closely the standard
(solar) flare scenario. 
 Time-resolved spectroscopy during the X-ray flare reveals, in addition to the time evolution of plasma temperature and emission measure, a temporary increase of electron density and elemental abundances, and a loop that extends in the corona  by $13$\,\% of the stellar radius ($4 \cdot 10^9$\,cm).    Independent estimates of the footprint area of the flare from TESS and {\em XMM-Newton} data 
 suggest a high temperature of the optical flare ($25000$\,K), but we consider more likely that the optical and X-ray flare areas represent physically distinct regions in the atmosphere of AD\,Leo. 
}

   \maketitle
%

\section{Introduction}\label{sect:intro}
Contemporaneous multi-wavelength data during flares is 
crucial for understanding the physics of the flare process.
Prominent characteristics expected from the standard solar flare scenario (the so-called CSHKP model) include a time lag between diagnostics for the impulsive and the gradual phase 
\citep[see e.g.][]{Benz02.0} and a relation between non-thermal and thermal emission, the so-called Neupert effect \citep{Neupert68.0} that was occasionally observed in large stellar flares as a correspondence between the profile of radio luminosity and the time-derivative of the X-ray luminosity \citep[e.g.][]{Guedel02.0, Osten07.0}.  

Stellar flares are also 
known to be a major driver for the evolution of planet atmospheres \citep[e.g.][]{Owen20.0}. 
However, the majority of exoplanet systems are too distant for a detailed characterization 
of the high-energy (X-ray and UV) emission of the host star. Therefore, models for the effects of stellar 
irradiation on planets are often based on the observed properties of individual well-known 
flare stars. Most notably a prototypical flare on \adleo, the so-called `Great Flare of 1985'  \citep{Hawley91.0},  
has been the basis for seminal work on the impact of stellar variability on 
planetary chemistry \citep{Segura10.0, Tilley19.0}.

\adleo is an M-type main-sequence star (SpT M3.5) at a distance of $4.966 \pm 0.002$\,pc \citep{GaiaEDR3}. 
It has an effective temperature of  $T_{\rm eff} = 3414 \pm 100$\,K and a  radius of $R_* = 0.426 \pm  0.049$\,$R_\odot$  \citep{Houdebine16.0}.  
Its rotation period of $2.23^{+0.36}_{-0.27}$\,d   \citep[measured on the MOST lightcurve,][]{HuntWalker12.0} and its 
X-ray luminosity of $\log{L_{\rm x}}\,[\rm erg/s] = 28.8$ \citep{Robrade05.2} place the star in the 
saturated regime of the rotation-activity relation where the X-ray emission level does not 
depend on rotation. 
For M dwarfs
the spin-down and associated diminishing of activity last up 
to $\sim$\,1\,Gyr \citep{Magaudda20.0, Johnstone21.0}. 
Therefore it is difficult to place an age constraint 
on AD\,Leo.  
While it appears as a typical M dwarf star based on its rotation and X-ray 
emission level, it is certainly one of the most studied stars in the northern hemisphere. 

Thanks to its extraordinary 
brightness originating in its favorable sky position \adleo has become the prototype 
for M-type dwarf stars which account for $\sim 75$\,\% of the stars in the Galaxy 
\citep[e.g.,][]{Chabrier01.0}.
The number of planets known to orbit such stars has been estimated to be very high, especially 
for low-mass planets that are the most suitable candidates for being habitable 
\citep{Dressing13.0, Sabotta21.0}. An entire space mission is dedicated to the 
discovery of planets around M dwarfs, the {\em Transiting Exoplanet Survey Satellite} 
\citep[TESS;][]{Ricker14.0}.  

Using the observed UV spectrum of the 1985 \adleo flare  \cite{Venot16.0} simulated the effect of stellar flares onto exoplanet spectra and found that the stellar flare radiation can induce irreversible changes in the chemical composition of hot planets; see also \cite{Chen21.0}.  The X-ray component of the flare was not considered in these studies as there was no contemporaneous data taken in that energy band for the 1985 flare of AD\,Leo. However, X-ray photons penetrate deeper into the planet atmosphere and have been shown to drive ionization and chemistry in gaseous exoplanets at layers inaccessible to UV radiation \citep[see][and references therein]{Locci22.0}.
\adleo itself has been reported from a radial velocity 
study to host a hot Jupiter planet \citep{Tuomi18.0}, but the signal was later attributed to 
stellar activity \citep{Carleo20.0}.

The studies discussed above have pointed out the importance of considering that flares are repetitive events. However, flare rates are poorly constrained in the crucial high-energy XUV band.  
While the planet transit search satellites {\em Kepler} and TESS have provided 
high-quality optical light curves for numerous flare stars, no instruments are available
that are suitable for a systematic monitoring of X-ray and UV flares. This hampers also the full characterization
of the dynamics and energy output of flares which requires their simultaneous detection in 
different wavebands. 
That this is a 
difficult task, 
can be 
assessed e.g. from the study of \cite{Namekata20.0} dedicated to optical and X-ray
monitoring of \adleo where in $8.5$\,nights of observations only one small flare was
observed jointly in X-rays with NICER \citep{Arzoumanian14.0} and optical instruments.


This article is dedicated to the characterization of a superflare\footnote{A superflare is commonly defined as an event with a radiative energy release of at least $10^{33}$\,erg;  \cite{Schaefer00.0}.}
on \adleo that was observed
during a recent pointing of the X-ray satellite {\em XMM-Newton}, and for which we detected the
optical counterpart in the TESS light curve. 




\section{Analysis of the superflare}\label{sect:analysis}

On Nov 18/19, 2021, AD\,Leo was observed for $86$\,ksec with {\em XMM-Newton} through 
Director's Discretionary Time. 
Due to the proximity of the bright star $\gamma$\,Leo ($V=1.98$\,mag),  located about $4.8^{\prime}$ south east of AD\,Leo, the X-ray pointing was performed in {\sc small window mode} for the prime instrument EPIC and the Optical Monitor 
had to
be kept in closed position. 
The data reduction for EPIC/pn has been carried out with a standard procedure that is described in Appendix~\ref{app:xmm_extraction}. 

The roughly one-day long {\em XMM-Newton} observation is fully covered with TESS's Sector\,45 that covered the time span from Nov 6  to Dec 2, 2021. The most evident feature is a huge flare towards the end of the {\em XMM-Newton} exposure which has a counterpart in the TESS data.
The optical flare is barely visible in the TESS PDCSAP light curve and the averaged target pixel file (TPF) available at the Barbara A. Mikulski Archive for Space Telescopes (MAST) Portal, due to the high noise level induced by $\gamma$\,Leo. 
We, therefore, had to perform a customized data reduction (explained in Appendix~\ref{app:average_tpf}) to reduce the noise.

\begin{figure}[t]
\begin{center}
\includegraphics[width=8.8cm]{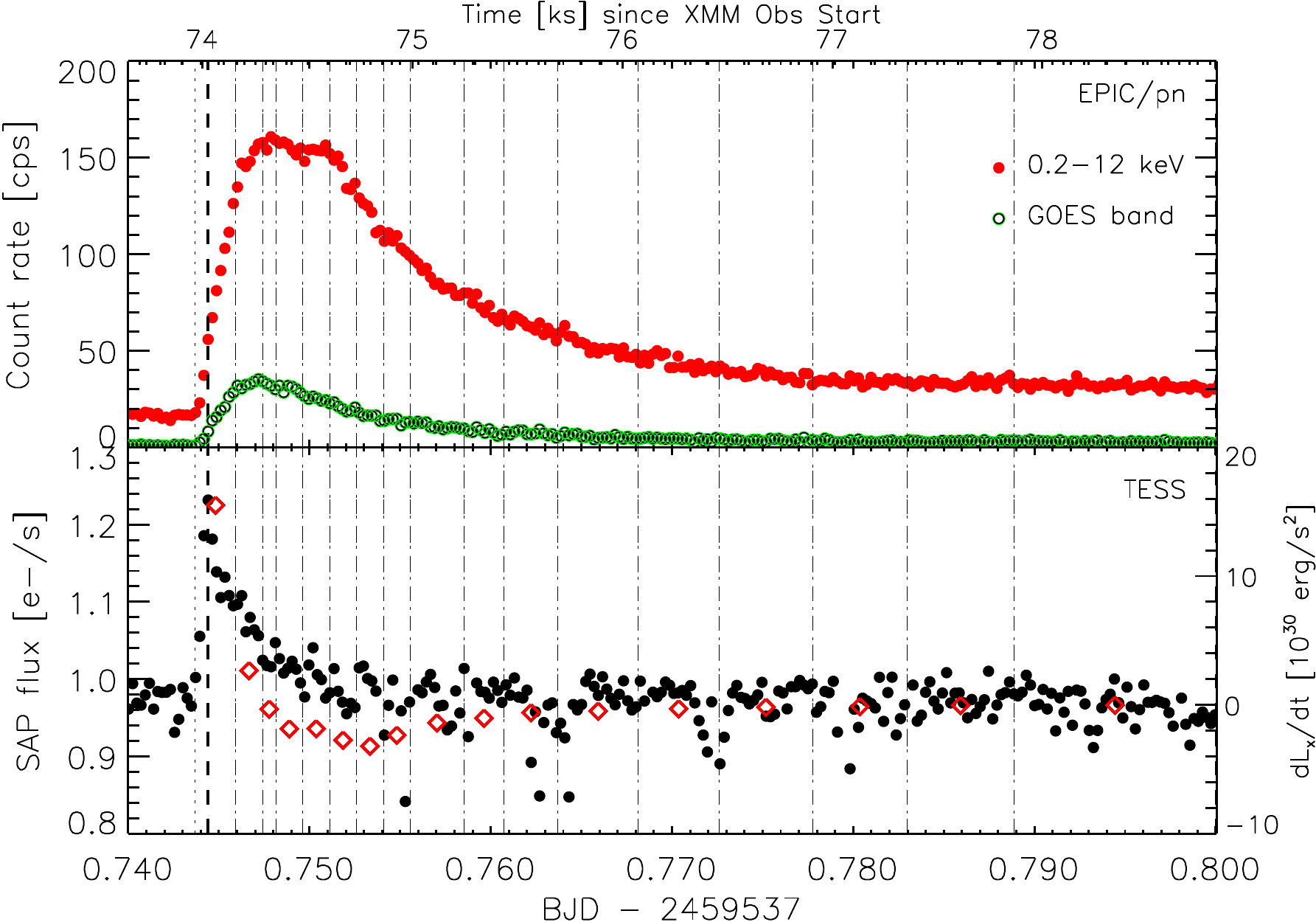}

\caption{Light curves of the superflare: {\em Top panel} - {\em XMM-Newton} EPIC/pn count rate in $0.2-10$\,keV (red) and in the GOES band ($1.5 - 12.4$\,keV; green), with the binning adapted to the 
TESS light curve in {\sc fast} cadence ($20$\,s) shown in the {\it bottom panel} (black circles), where it is overlaid with the time derivative of the X-ray luminosity (open diamonds). The thick dashed vertical line marks the time of the peak of the flare flux in the {\em TESS} light curve, and the dash-dotted lines mark the boundaries of the intervals used for the time-resolved spectral analysis of the X-ray data. For a better  display of the main flare phase a cut has been set to the abscissa values such that the tail of the flare is not shown.}
\label{fig:lcs_superflare}
\end{center}
\end{figure}

In Fig.~\ref{fig:lcs_superflare} we present the  simultaneous X-ray and optical light curves of the  superflare together  with the time derivative of the X-ray luminosity. In the remainder of this section we describe how we extracted physical parameters from the X-ray and optical data of the flare.
 
%

\subsection{X-ray data}\label{subsect:analysis_xmm}

The time profile of the X-ray flare is shown in the top panel of  Fig.~\ref{fig:lcs_superflare} for the broad {\em XMM-Newton} energy band ($0.2-12$\,keV) and for the GOES band ($1-8$\,\AA). In the {\em XMM-Newton} broad band it displays a roughly linear (about $340$\,s long) rise
phase, followed by a plateau (lasting about $400$\,s).
The subsequent decay 
can be described by an exponential phase with a time-scale of 
$\tau_{\rm exp} = 724 \pm 8$\,s 
and a linear phase\footnote{We determined the transition between the exponential and the linear phase by fitting a decaying  exponential function to the decreasing part of the light curve starting with the first four bins after the end of the peak phase and successively adding data points until the minimum of   $\chi^2_{\rm red}$ was reached.} 
that lasts for the remaining roughly $6$\,ks until the end of the observation. 
At the end of the observation the count rate was still above the pre-flare level suggesting that the underlying quiescent corona slightly changed  during the flare. 
Despite this prolonged tail 
 the initial fast decay of the X-ray light curve indicates a short duration event, hence likely occurring in a compact coronal structure. 


\subsubsection{X-ray spectral analysis}\label{subsubsect:analysis_xray_spectra}

We have divided the flare from the start of its rise (2021-11-19 05:48:50.816 UTC) to the end of the exponential phase (2021-11-19 07:08:50.816 UTC) into time intervals of about $10000$~EPIC/pn counts each. The $17$ time bins obtained this way have different duration but roughly the same photon statistics. 
We extracted an EPIC/pn spectrum from each of the $17$ intervals and subtracted the out-of-time events. 

Our goal to constrain the physical conditions in the corona of \adleo during the flare requires an accurate assessment of the underlying quiescent, i.e. non-flaring, X-ray emission. To this end, we have identified all flare-free parts of the 
EPIC/pn light curve and combined them into one spectrum. 
We used this quiescent spectrum 
as background for the study of the spectral evolution
during the X-ray flare. This way we obtained a series of ‘flare-only‘ spectra. 


The spectrum of each of the $17$ individual time slices of the flare was fitted in XSPEC v 12.11.1 \citep {Arnaud96.0} with a two-temperature {\sc vapec} model. To avoid an excessive number of free parameters we 
tied the abundances of all elements ($X$) to that of iron according to the ratio $A_{\rm X}/A_{\rm Fe}$ determined for the quiescent state. 
The remaining elements considered in {\sc vapec} (He, Ca, Al, Ni), which have no significant emission lines in the spectral range examined, were fixed to the solar values. 
Since low-resolution X-ray spectra are notoriously affected by degeneracies between abundances and emission measure ($EM$) we 
derived 
the abundances of the quiescent emission of \adleo 
from the high-resolution RGS spectrum,  
that 
was extracted on the same time intervals as the quiescent EPIC/pn spectrum.  
We used the APED database \citep{Smith01.0} and the updated solar abundance table of \cite{Asplund09.0}.  
The full emission measure distribution analysis of the RGS data  
will be explained elsewhere.  
In Table~\ref{tab:rgs_abund} we report the abundances 
obtained from the quiescent RGS spectrum, since their ratios were used to restrict the spectral model for the flare state as explained above. 
Free fit parameters for the $17$ EPIC/pn flare spectra were, thus, the two temperatures and two emission measures and the abundance of Fe. Best fit results are listed in Table~\ref{tab:EPICfitres}, where we also report, for each time interval, the $EM$-weighted average temperature and the total $EM$.

\begin{figure}[t]
\begin{center}
\includegraphics[width=0.5\textwidth]{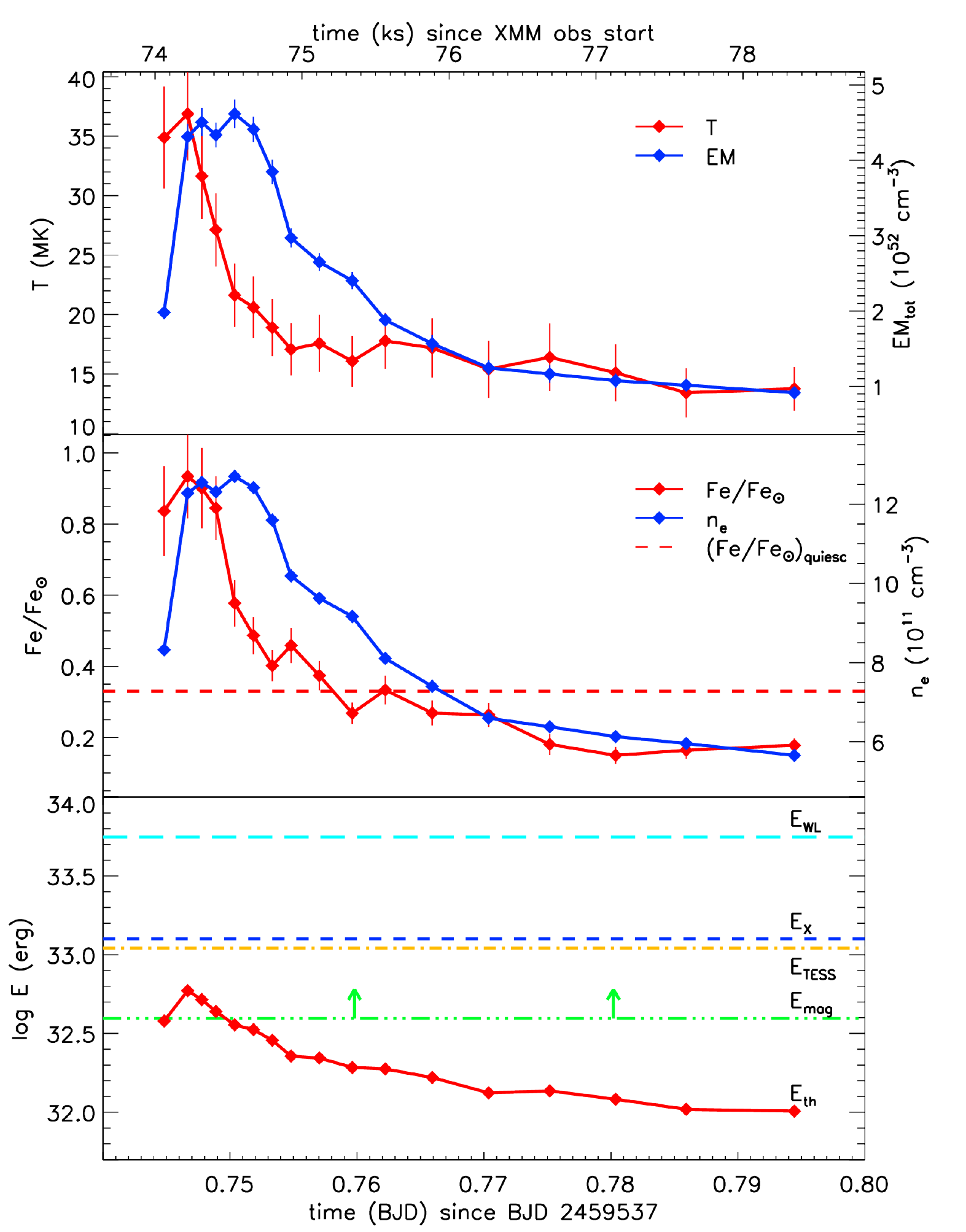}
\caption{{\it Upper panel}: Time evolution of the flaring plasma temperature and emission measure. {\it Middle Panel}: Time evolution of flaring plasma abundances and electron density. The abundances of all elements are tied to the Fe abundance using the ratios determined for the quiescent state from the RGS spectrum that are listed in Table~\ref{tab:rgs_abund}. {\it Lower Panel}: Time evolution of the thermal energy in the flaring plasma, compared with total radiated energy in the X-ray and optical bands.} 
\label{fig:flare_results}
\end{center}
\end{figure}

The resulting time evolution of the  spectral parameters during the flare  ($EM$-weighted average temperature $T$, sum of the $EM$ of the two components $EM_{\rm tot}$, and iron abundance) 
 is shown in Fig.~\ref{fig:flare_results}. By definition of the spectral model the  variation of Fe  includes the variation of the abundances of the other elements. It can be noted from Fig.~\ref{fig:flare_results} that in the tail of the exponential decay phase the flare Fe abundance falls below the quiescent value. This likely indicates a change of the underlying quiescent corona that is manifest also in the elevated count rate after the flare (see Fig.~\ref{fig:lcs_superflare}), and that will be investigated in a future work.  
%

The top panel of Fig.~\ref{fig:flare_results} shows that the
 peak temperature 
 is reached before the peak of the $EM$, as expected if the enhanced X-ray radiation is caused by a heating event. Secondly, the temperature rapidly drops at nearly constant $EM$ (the plateau at the maximum in the light curve). 
 At the onset of the decrease of the $EM$, the temperature has already decayed to about half its peak value. 

Integrating over the fluxes in the $17$ time slices we determined the flare energy. For the GOES band we found 
$(4.30 \pm 0.05)  \cdot 10^{32}$\,erg (see also Table~\ref{tab:flareparameters}) and for the {\em XMM-Newton} X-ray band ($0.2-12$\,keV) we found 
$E_{\rm F,XMM} = (1.26 \pm 0.01) \cdot 10^{33}$ \,erg.  
 

\subsubsection{Physical conditions of the X-ray flare}\label{subsubsect:analysis_xray_physics}

\begin{figure}
\begin{center}
\includegraphics[width=8.5cm]{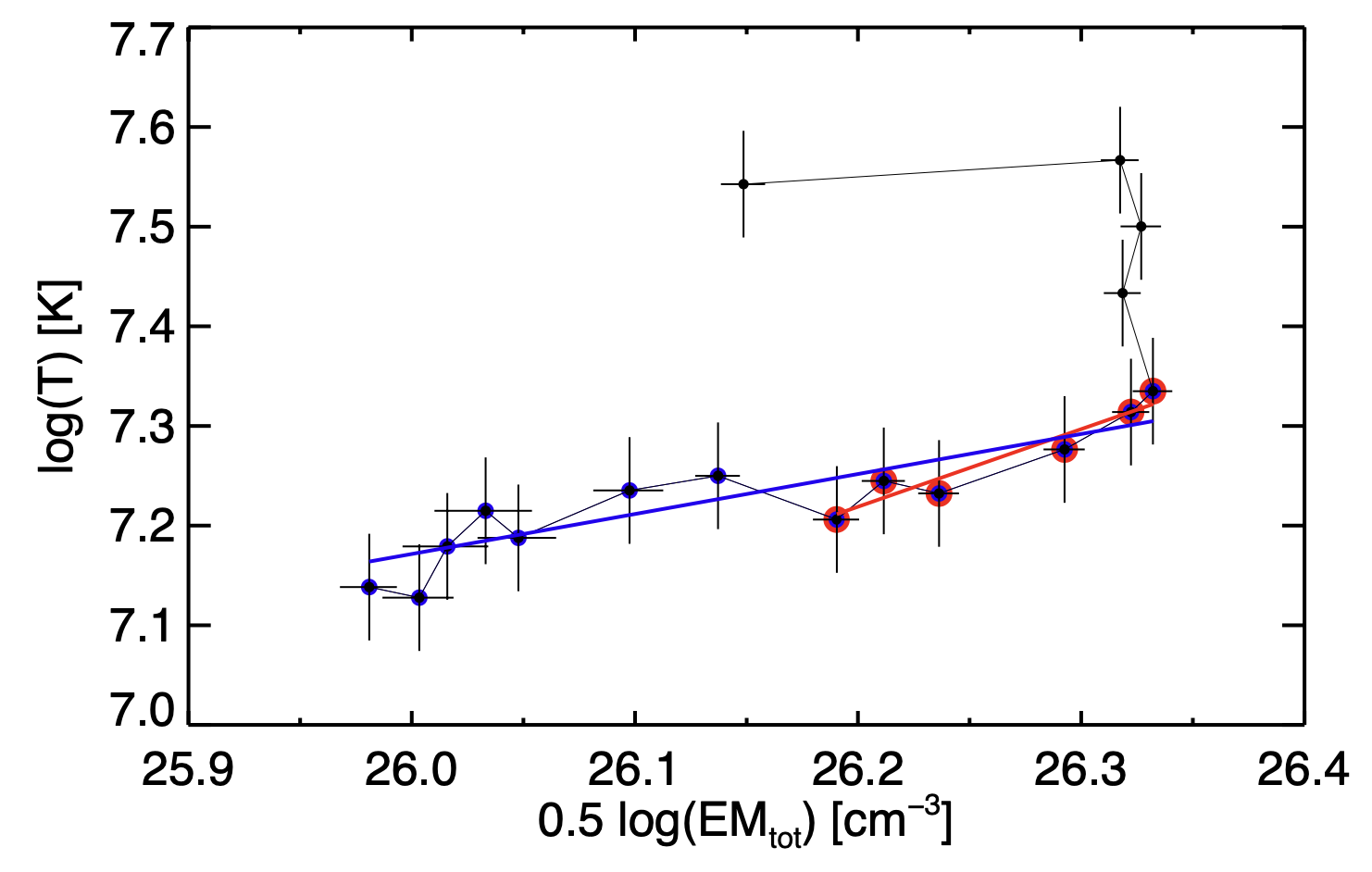}
\caption{Time evolution of the mean temperature and emission measure associated with the
flare. The red line represents a linear least-square fit to this first phase of the flare decay (when the flaring emission is still much higher than the slightly-variable background corona). The blue line is the fit to the whole exponential decay.} 
\label{fig:em_t_diagram}
\end{center}
\end{figure}

The time-resolved spectroscopy during the flare decay can be used to determine the semi-length, $L$, of the  flaring loop making use of the prescription of \cite{Reale97.0}, who have performed hydrodynamic simulations to predict the X-ray
spectral signature of decaying flare loops. Assuming that the flaring structure has a constant volume ($V$) with a uniform cross section ($S$) and that the loop has a half-torus shape (hence its volume is $V = 2 \cdot S \cdot L$), its semi-length can be inferred by inspecting the evolution of plasma temperature and density during the decay phase. The hypothesis of constant volume implies that the plasma density, $n$, is proportional to $\sqrt{EM}$.  
The slope, $\zeta$, of the trajectory traced by the flaring plasma in the $\log{T}$ versus $\log{n}$ space, shown in Fig.~\ref{fig:em_t_diagram}, allows us  then to infer the amount of heat released into the loop during the decay. 
Combining this with the observed exponential decay time inferred from the light curve 
($724\pm 8$ \,s)
and the temperature at the peak of the flare  ($\log{T_{\rm peak}}= 7.57 \pm 0.05$\,K) the loop semi-length can be estimated. 
We observe that after the first rapid temperature decay at constant $EM$ 
the $\log{T}$ versus $\log{n}$ evolution displays first a  
joint
decrease of both 
quantities, followed by a minor re-heating event. We infer the slope $\zeta$ considering only the first decay path\footnote{
In this first phase of the decay 
the flaring emission is still significantly higher than that of the background corona. Therefore, the inferred quantities are not significantly affected by the changes occurring in the underlying corona, that is not accounted for in our analysis. 
We note, however, that including the whole decay path in the fit (blue points and blue line in Fig.~\ref{fig:em_t_diagram}) yields a slope of $0.4 \pm 0.2$,  comprised within the errors in the value obtained for the reduced time-span.} (red points and red line in Fig.~\ref{fig:em_t_diagram}), obtaining $\zeta=0.8\pm0.6$.  
The loop semi-length obtained with 
the equations of \citet{Reale04.0}, where the procedure calibrated for the EPIC/pn detector is derived, 
is then $L \approx 4 \times 10^9$\,cm.
Note, that the equation that yields the loop length has been derived by \cite{Reale97.0} under a series of assumptions involving the loop geometry (see beginning of this section) and heating (exponentially decaying). Therefore, the value of $L$ and all other quantities we derive in the following from this parameter are order of magnitude estimates. 
%


If the plasma density is known, together with the loop semi-length, the volume and cross-section of the flaring loop can be determined 
from the assumptions on the geometry and the definition of the $EM$.
The electron density can be estimated from the high-resolution RGS spectrum, making use of the ratio between forbidden and intercombination lines of He-like triplets \citep{Gabriel1969}. To this end, we extracted an RGS spectrum  considering the entire exponential flare duration (that is the time span that encompasses all the $17$ intervals used for the EPIC/pn time-resolved analysis). 
Details on the RGS analysis will be provided in a forthcoming paper. 
For the purpose of the current work, 
we subtracted the RGS spectrum of the quiescent phase from the RGS flare spectrum, to 
inspect the strongest triplets, those of  \ion{O}{\sc vii} and \ion{Ne}{\sc ix}. 
The significant emission observed in the triplets for this net flare-phase RGS spectrum confirms that, even if their formation temperatures are quite low ($\sim2$ and $\sim4$\,MK respectively), the flaring plasma significantly 
contributes to the emission in 
these lines. The \ion{O}{\sc vii} and \ion{Ne}{\sc ix} triplets indicate electron densities $n_{\mathrm{e}}$ of $5^{+3}_{-2}\cdot10^{10}$ and $8^{+10}_{-7}\cdot10^{11}$\,cm$^{-3}$. 
For the evaluation of the geometric loop parameters we considered the $n_{\mathrm{e}}$ value of $8\times10^{11}$\,cm$^{-3}$ 
since the higher formation temperature  of \ion{Ne}{\sc ix} suggests that it probes the flaring plasma density better than the cooler \ion{O}{\sc vii}. Combining this value with the total $EM$ of 
$(1.83 \pm 0.03)  \cdot 10^{52}$\,cm$^{-3}$  
of the entire flare\footnote{This value 
derives from 
averaging $EM_{\rm tot}$ of the $17$ time intervals.} and with the loop semi-length, and assuming $n_{\mathrm H}/n_{\mathrm e}=0.83$  
(proper for typical coronal temperatures and chemical compositions), we derive for the flaring loop a volume of $3 \cdot 10^{28}$\,cm$^3$ and a cross section 
$S \approx 5 \cdot 10^{18}$\,cm$^2$.

Having estimated the loop volume, together with the detailed evolution of its 
$T$ and $EM_{\rm tot}$ during the flare, gives us the unique opportunity to probe the physical conditions of the flaring plasma during the flare's evolution. Firstly, we can infer the evolution of the flaring plasma density $n_{\mathrm e}$, obtained as $\sqrt{EM/(0.83V)}$  
(Fig.~\ref{fig:flare_results}, middle panel). We can also compute the evolution of the total thermal energy  $E_{\rm th}=(3/2)(n_{\mathrm e}+n_{\mathrm H})\,V\,k_{\rm B}T$ of the flaring plasma 
(Fig.~\ref{fig:flare_results}, lower panel). In addition, the knowledge of plasma density and temperature allows us to probe the pressure experienced by the flaring plasma, i.e. $P_{\rm gas}=(n_{\mathrm e}+n_{\mathrm H})k_{\rm B}T$. Since this plasma is magnetically confined, the highest value of $P_{\rm gas}$ provides us a lower limit for the magnetic pressure,  $P_{\rm mag, min} = B^{2}/(8\pi)$, which in turn implies a minimum magnetic field strength ($B$) of $500$\,G, and a minimum magnetic energy for the flaring loop of $E_{\rm mag, min} = 
P_{\rm mag, min} \cdot V =  4\times10^{32}$\,erg (also plotted in the lower panel of Fig.~\ref{fig:flare_results}).

\subsection{Optical data}\label{subsect:analysis_tess}

We searched for the rotational signal and for flares in the TESS light curve following our previous work,  \citet{2016MNRAS.463.1844S} and  \citet{2020A&A...637A..22R} on data from the \textit{K2} mission, and  \citet[][period search adapted for TESS data]{Magaudda22.0} and \citet[][flare search on TESS data]{Stelzer22.0}. 

\subsubsection{Analysis of the rotation signal}
\label{subsubsect:analysis_tess_rotmod}



A 
detailed description and a graphical illustration of our period search on \adleo can be found in Appendix~\ref{app:tess_rotation}. 
We found a rotation period of $2.194 \pm 0.004$ which is consistent with the period found by \citet{HuntWalker12.0}. 
The half-amplitude of the rotation signal is $0.00217 \pm 0.00007$. 
We estimated the spot coverage of AD\,Leo using the relations given by \citet{2019ApJ...876...58N}. With their Eq.~4, which they deduced from \citet{2005LRSP....2....8B}, the spot temperature, $T_{\rm spot}$, was computed from AD\,Leo's effective temperature 
With the computed value of 
$T_{\mathrm{spot}}=2955$\,K 
we found with Eq.~3 of \citet{2019ApJ...876...58N}  a spot filling factor of $A_{\mathrm{spot}}/A_{\mathrm{star}} \approx 1.0$\,\%, where $A_{\rm spot}$ is the spotted area and $A_{\rm star}$ the total surface area of the star. 
 
 \subsubsection{Flare analysis}\label{subsubsect:analysis_tess_flare}


We validated four flare candidates in the full light curves, among  which the X-ray superflare that was clearly detected in the TESS light curve (see bottom panel of Fig.~\ref{fig:lcs_superflare}). 
In Table~\ref{tab:flareparameters} we provide all relevant 
flare properties for the superflare determined by our algorithm, namely  
the duration ($\Delta t$, the time between the first and last flare point), the relative peak flare amplitude ($A_{\mathrm{peak}}$,  the continuum flux level subtracted from the flux of the peak), the absolute peak flare amplitude ($\Delta L$, multiplying the $A_{\mathrm{peak}}$ by the quiescent stellar luminosity) and the equivalent duration ($ED$, integral under the flare). 
Following \citet{Davenport16.0}, we calculated the  
flare energy, $E_{\rm F}$, by multiplying the $ED$ with the quiescent stellar luminosity 
 of \adleo in the TESS band, that we determined from the TESS magnitude of \adleo ($T = 7.036$) to 
 $L_{\rm qui, T}=2.4 \cdot 10^{31}$\,erg/s.
This value is consistent within $5$\,\% with the luminosity 
that we obtain if 
we use AD\,Leo's effective temperature and radius, and  integrate the blackbody function taking into account the TESS filter transmission. 
Assuming for the optical flare a black-body emission at a constant temperature of $9000$\,K \citep[as typically observed for solar WLFs,][]{ Kretzschmar11.0},  the emitting area (assumed to be variable) can be constrained to match the observed amplitude of the TESS light curve, $L_{\rm F,T}(t)/L_{\rm qui,T}$. 
We found a maximum value for the area of $8.4 \cdot 10^{19}\,{\rm cm^2}$. Multiplying this by the flare surface flux 
yields the bolometric flare luminosity ($L_{\rm F, bol} = 3.1 \cdot 10^{31}$\,erg/s),  
and integration over the flare light curve gives the 
bolometric energy radiated by the flare 
($E_{\rm F,bol} = (5.57 \pm 0.03) \cdot 10^{33}$\,erg). 

 \begin{table}
  \begin{center}
    \caption{Parameters of the superflare extracted from the optical and X-ray light curves.}
    \label{tab:flareparameters}
    \begin{tabular}{lcc} \hline
      Parameter [unit] & TESS & {\em XMM-Newton}$^{(a)}$ \\ \hline
      $t_{\rm peak}$ [BJD-2459537] & 0.7444 & 0.7479 \\
      $\Delta t$ [min]  & $6.67 \pm 0.67$   & $\approx 82.0$   \\
      $A_{\rm peak}$ & $0.259 \pm 0.006$ & - \\
      $\log{\Delta L}$ [erg/s] & $30.79 \pm 0.03$ & 
      $29.66 \pm 0.02$ \\
      ED [s] & $44.9 \pm 0.3$ & - \\
      $E_{\rm F}$ [erg] & $(1.06 \pm 0.05) \cdot 10^{33}$ & 
      $(4.30\pm 0.05) \cdot 10^{32}$\\ \hline
\end{tabular}
\tablefoot{$^{(a)}$ Values are for the GOES band from the start of the rise phase to the end of the exponential phase.}
\end{center}
\end{table}

\section{Discussion}\label{sect:discussion}


The Nov 2021 flare on \adleo 
has an energy above the canonical threshold for a superflare, $10^{33}$\,erg, in both the TESS and the {\em XMM-Newton} band. It was a factor $30$ stronger than the largest solar flare observed to date, the Sep 1859 Carrington event, which was of GOES class X45
\citep{Hudson21.0}, while for the \adleo superflare we measure a peak flux in
the $1-8$\,\AA~ band of 
$(1.38 \pm 0.03)\cdot 10^{-10}\,{\rm erg/cm^2/s}$, 
corresponding to an X1445 event on the GOES flux scale.  
Yet, it is a small event when compared with the largest superflares reported from main-sequence stars in the optical band  \citep[e.g.][]{Schaefer00.0, Maehara12.0}. In X-rays, on the other hand, observations of giant flares have mostly been limited to pre-main sequence objects or interacting binaries  \citep[e.g.][]{Preibisch95.0, Grosso97.0, Pandey12.0, Getman21.0}. 


AD\,Leo's Nov 2021 event 
displays a time-profile similar to the one observed in a standard solar flare,
where optical emission, 
associated to the energy deposited by non-thermal high-energy particles in the lower layers of the stellar atmosphere,
precedes the X-ray emission peak, 
due to the subsequent chromospheric evaporation
\cite[see e.g.][]{CastellanosDuran20.0}.
The brightness peak in the optical is observed about $300$\,s before the X-ray maximum in the GOES band. The optical
light curve is strongly peaked while the X-ray maximum is a plateau that makes a transition into
an exponential decay followed by a slow linear decrease. 
The chromospheric evaporation scenario  \citep[see e.g.][]{Benz17.0} is further corroborated by the increase of density and elemental abundances observed during the flare, which provide also new constraints on metal depletion 
in coronal plasma. 
The higher abundance of the flaring ($\mathrm{Fe/Fe_{\odot}}\sim0.9$), compared to the quiescent plasma  ($\mathrm{Fe/Fe_{\odot}}\sim0.3$), clearly proves that the quiescent corona is metal depleted with respect to the chromospheric material, that manifests its higher metallicity in the X-rays during the initial phases of the flare. For the first time we constrained the time scales of coronal metal depletion, through the rapid decrease (in a few $100$\,s) of the elemental abundances after the chromospheric evaporation event.


In absence of data in the radio band, the WLF seen with TESS, can be taken as a proxy for the non-thermal component, because in the standard flare scenario \citep[e.g.][]{Benz17.0} it is produced by the bombardment of  lower atmospheric layers with (non-thermal)  electrons from the magnetic reconnection site. 
We have calculated  the evolution of the time derivative of $L_{\rm x}$,  
shown as open diamonds in the bottom panel of Fig.~\ref{fig:lcs_superflare},  from the fluxes measured in the $17$ time bins representing the exponential flare phase. The behavior follows that of the WLF demonstrating the presence of the Neupert effect.

The TESS and {\em XMM-Newton} data have provided independent estimates for the surface coverage of the flare footprint. 
The time-averaged area of the optical flare, determined from the changing  amplitude of the TESS light curve, is 
larger than the X-ray based area by about a factor seven. 
A temperature of $25000$\,K for the optical flare would reconcile this discrepancy. However, it is likely that the two measurements probe different emitting regions. The X-ray emission, produced by optically thin plasma, provides  us with the cross-section of the coronal part of the loop, while 
the optical emission originates from the optically-thick lower layers of the flaring structure. Hence, the area inferred from the optical flare  possibly embraces both the horizontal extent and vertical structuring of the flaring loop foot-points. 
Given that the TESS light curve displays a fairly regular sine-like rotational modulation (Fig.~\ref{fig:tess_rotation}), at the epoch of the observation, AD\,Leo's photosphere was possibly dominated by a single spot group. 
The flare took place when this spotted part of the photosphere was located near the limb\footnote{The flare occurred at phase $0.78$  of the rotational signal, with phase $0.0$ corresponding to the minimum flux. The low inclination of AD\,Leo ($i\sim15^{\circ}$), together with the unconstrained spot latitude, does not allow us to guess more precisely the spot position.}. If the flare was, indeed, spatially connected to the spot, as is typically seen in solar flares \citep[e.g.][]{Toriumi19.0}, we had a lateral view onto the loop structure. We speculate that in such a geometry one can account for the high observed flare area inferred from the TESS data if the optical flare region had a significant vertical extent.

\begin{figure}
     \centering
     \includegraphics[width=9cm]{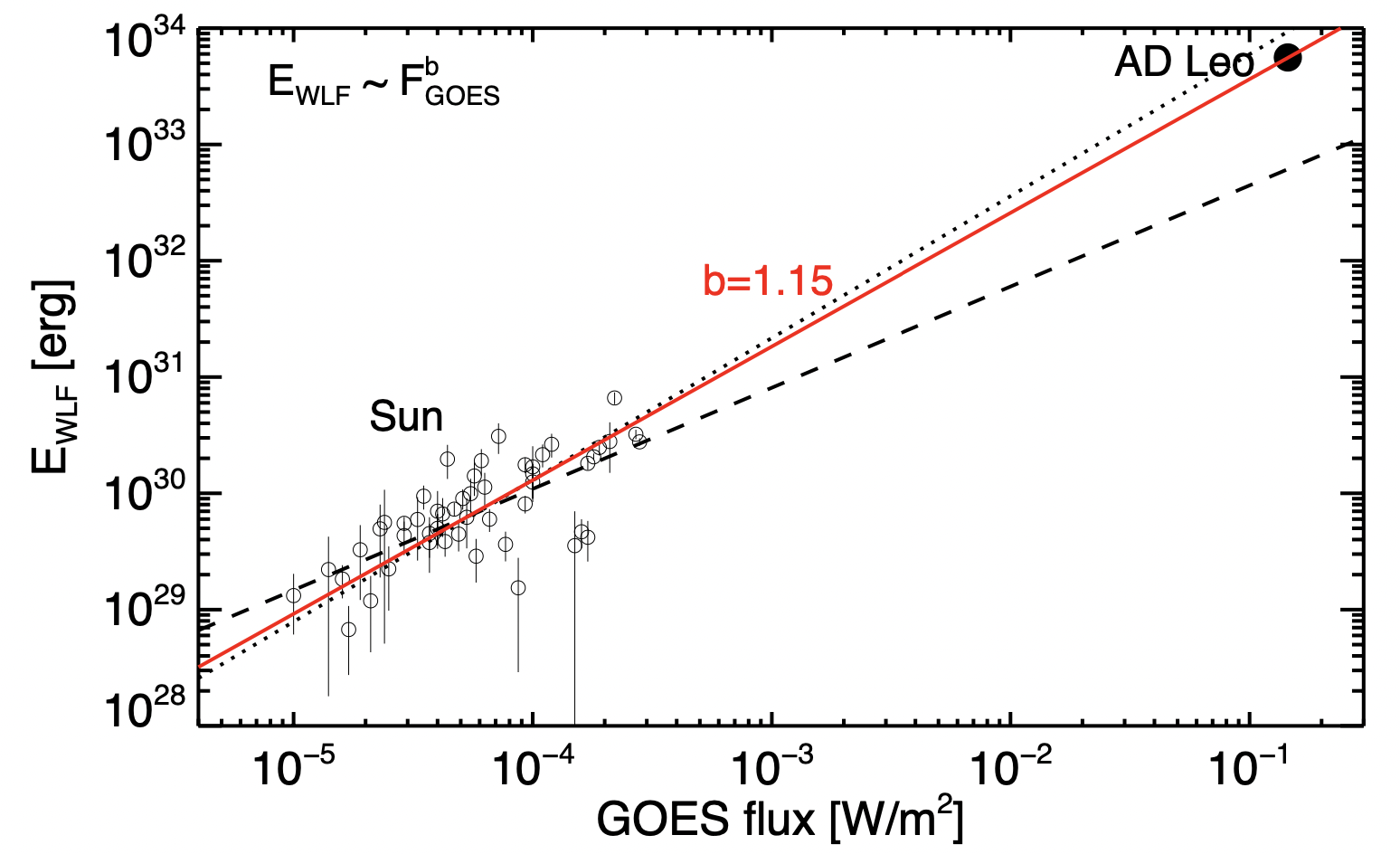}
     \caption{Empirical relation between WLF energy and GOES flux for solar flares and power-law fits performed by \cite{Namekata17.0} on them using a linear regression method and a linear regression bisector method (solid and dashed lines). The red line is our fit including the \adleo superflare which yields in the double logarithmic form ($\log{E_{\rm WLF}} = a + b \cdot \log{F_{\rm GOES}}$) a slope $b  = 1.150 \pm 0.005$ and an axis-offset of $a = 34.711 \pm 0.007$, where the uncertainties are the standard deviations of the fit.  
     }
     \label{fig:namekata17_and_adleo}
 \end{figure}

 Fig.~\ref{fig:namekata17_and_adleo} shows that the AD\,Leo superflare is consistent with the extrapolation of the power law relation between white-light energy and GOES peak flux  derived for the solar flare sample of \cite{Namekata17.0}. The inclusion of \adleo increases the range of values by several orders of magnitude, and allows - contrary to Namekata et al.'s  study based solely on solar flares - to constrain the power-law relation.  
 %
 


In the empirical relation between flare duration ($t_{\rm F}$) and energy ($E_{\rm F}$) the \adleo WLF is placed among the smallest {\em Kepler} superflares on  solar-type (that is G-type main-sequence)  stars observed with similar  cadence, i.e. {\em Kepler} one-minute light curves  \citep{Namekata17.0}. 
Theoretical scaling laws presented by these authors predict for given $t_{\rm F}$ and $E_{\rm F}$ the magnetic field strength and coronal loop length.  
For AD\,Leo  
$B \sim 100-200$\,G  is found, somewhat smaller than the value we 
measured 
from the X-ray data ($B_{\rm min} \approx 500$\,G). In fact, \cite{Namekata17.0} showed by comparison to resolved solar flares that the scaling laws under-predict the field strength.
The loop length obtained for AD\,Leo's superflare from the scaling laws is $\approx 10^{10}$\,cm, about a factor two larger than the value derived from the X-ray analysis. 

A parameter that is of utmost importance for evaluating the impact of stellar flares on planets is the energetic particle flux that reaches the planet \citep[see e.g.][]{Tilley19.0}. For the Sun-Earth system calibrations 
between the flux of protons with energy $> 10$\,MeV ($I_{\rm p}$)  and the X-ray flux in the GOES band at the flare peak were recently updated by \cite{Herbst19.0}. We estimate for the X-ray superflare of \adleo a proton flux of 
$I_{\rm p} \sim 0.1 \cdot 10^6\,{\rm cm^{-2} s^{-1} sr^{-1}}$ (from Eq.~4 of \cite{Herbst19.0}) and 
$I_{\rm p} \sim 20.1  \cdot 10^6\,{\rm cm^{-2} s^{-1} sr^{-1}}$ 
(from their Eq.~5). An even larger possible range is obtained if the uncertainties of the solar empirical relations are folded in. \cite{Herbst19.0} present several stellar flares overlaid on the extrapolated solar relation between $I_{\rm p}$ and GOES flux, including {\em Kepler} superflares, and UV events on some benchmark M dwarfs. However, we stress that none of these events has an actual measurement of the X-ray peak flux, and the GOES class for all of them has been estimated using empirical relations to transform observed fluxes at lower wavelengths to the X-ray band introducing significant additional uncertainties.

\vspace*{0.5cm}









\section{Conclusions}

The high-cadence simultaneous coverage throughout the full event in the optical and X-ray band makes this superflare on \adleo a special calibrator for stellar flare physics and the stellar input to exoplanet atmospheres. 
Having sufficient signal at high energies to evaluate the flare energetics in the GOES band provides the rare possibility to quantify 
the relation between stellar X-ray superflares and  the much larger data base of solar flares. The \adleo flare exceeds the largest solar flare, the Carrington event, by a factor $30$ in peak X-ray flux, and by a factor $14$ in energy. 
Stellar flares with energies up to about $10^{37}$\,erg were reported in the literature. However, in most cases the radiative output in the X-ray band was estimated from empirical relations with optical or UV flare diagnostics that are subject to order of magnitude uncertainties. Parameters 
inferred in such an indirect way 
are correspondingly ill determined. 
With its  simultaneous WLF and GOES band measurements we could verify that the energetics of the \adleo flare, and therefore likely that of other stellar superflares, constitute a scaled up version of solar flares,
and we have derived the soft proton flux expected to be associated with the event that may serve for future exoplanet studies. 
\begin{acknowledgements}
We wish to thank the anonymous referee.  M.Caramazza is supported by the Bundesministerium für Wirtschaft und Energie through the Deutsches Zentrum für Luft- und Raumfahrt e.V. (DLR) under grant number FKZ 50 OR 2105. 
This research made use of observations obtained with {\em XMM-Newton}, an ESA science mission with instruments and contributions directly funded by ESA Member States and NASA. 
This paper includes data collected by the TESS mission, which are publicly available from the Mikulski Archive for Space Telescopes (MAST). Funding for the TESS mission is provided by NASA’s Science Mission directorate.
\end{acknowledgements}

\bibliographystyle{aa} 
\bibliography{adleo_superflare}

\newpage
\appendix
 
 \section{XMM-Newton EPIC/pn data extraction}\label{app:xmm_extraction}

We have analysed
the {\em XMM-Newton} observation using the 
Science Analysis Software (SAS) version 19.1.0 developed for the satellite. 
By examining the high energy events ($\ge$ $10$\,keV) across the full EPIC/pn detector, which are representative for the overall background, we have verified that 
the observation is not seriously affected by solar particle background. 
We filtered the data for pixel patterns ({\sc 0 $\leq$ pattern $\leq$ 12}), 
quality flag ({\sc flag = 0}) and events channels ({\sc 200 $\leq$ PI $\leq$ 15000}).   
Source detection was performed in three energy bands: 
$0.2-0.5$\,keV (S), $0.5-1.0$\,keV (M), and $1.0-2.0$\,keV (H), after having removed the out-of-time events.
For the spectral and temporal analysis we allowed only pixel patterns with {\sc flag $\leq$ 4}.
We defined a circular photon extraction region with radius of $30^{\prime\prime}$ centered on the EPIC/pn source position. 
The background was extracted from an adjacent circular region with radius of 
$45^{\prime\prime}$. 
The background subtraction of the light curve was carried out with the SAS task 
{\sc epiclccorr} which also makes corrections for instrumental
effects. We then barycentric corrected the photon arrival times
 using the SAS tool {\sc barycen}.

 \section{TESS data reduction}\label{app:average_tpf}


Here we show how we obtained a light curve with a lower noise level
than the PDCSAP light curve by removing the contribution of $\gamma$\,Leo from the flux in the pixels that we have identified as the most contaminated ones. 
We analysed the {\sc fast cadence} TESS light curve to obtain the best possible time-resolution. 

\begin{figure}[h]
\begin{center}
\includegraphics[width=8.0cm]{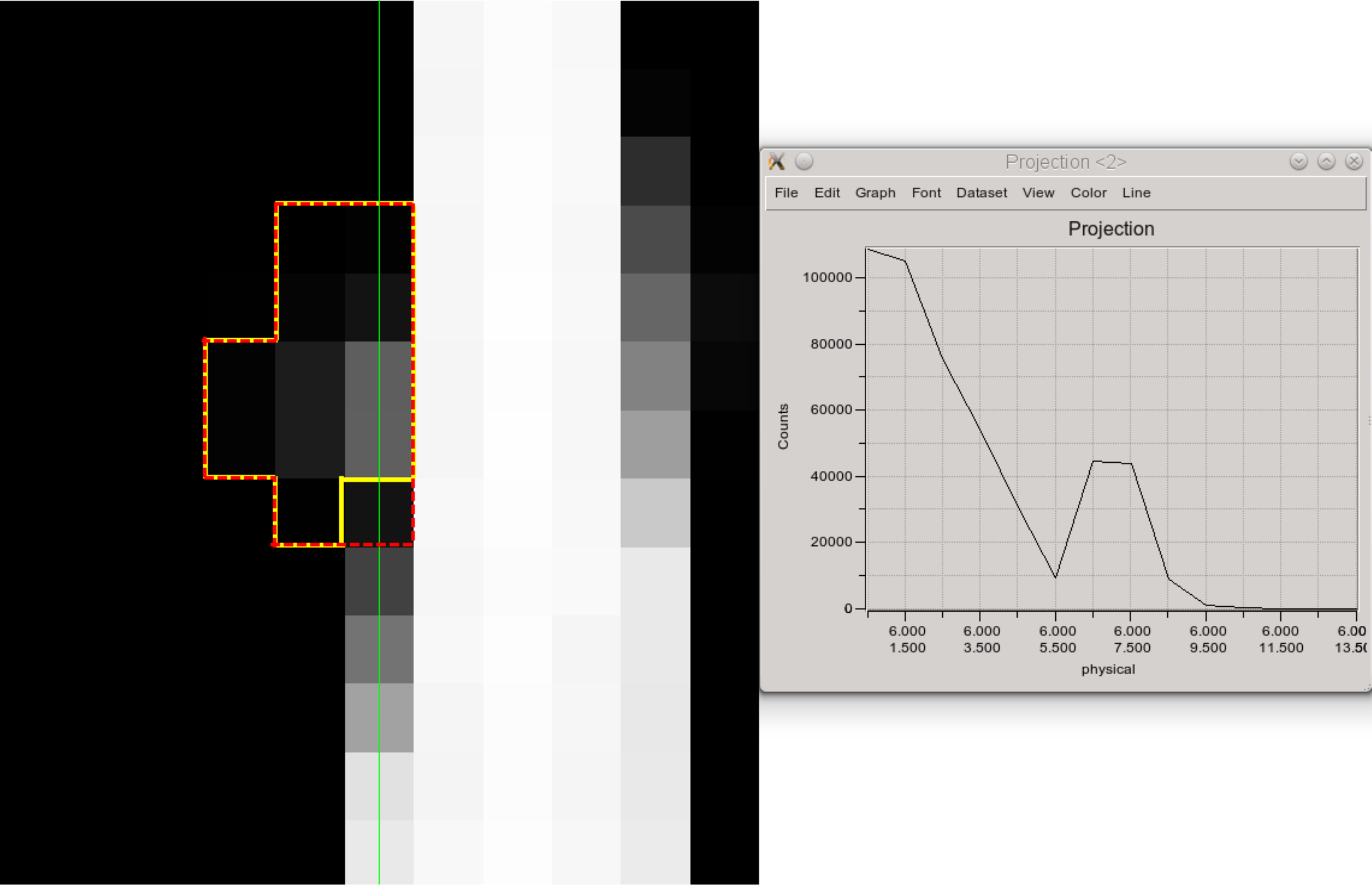}
\caption{Single frame of the \textit{TESS} target pixel file with the pipeline mask as red dashed line. The ``bleed trail'' of $\gamma$\,Leo contaminates the flux of AD\,Leo. The graph on the right shows the projection of the pixel column with the strongest contamination (given as the green line in the TPF). We applied our own light curve extraction using a customized mask  (yellow solid line) and corrected the flux level of the second strongest contaminated pixel (see text).}
\label{fig:TPF_masks}
\end{center}
\end{figure}

An evaluation of the individual frames of the TPF showed that the contamination is strongest close to the ``bleed trail'' of the saturated star $\gamma$\,Leo which extends to the pixel column of the TESS pipeline mask for AD\,Leo. 
We examined the contamination of the TPF by $\gamma$\,Leo  monitoring the flux level of all pixels in that column for each frame of the TPF. 
Based on this inspection 
we removed the flux of the most contaminated pixel (the lower right pixel of the pipeline mask; 
see Fig.~\ref{fig:TPF_masks})
from the light curve extraction. As a second step we fitted a Gaussian to the flux in that column and removed the flux of $\gamma$\,Leo from the second most contaminated pixel, the pixel above the most contaminated one. With this procedure we obtained a light curve with a lower noise level, decreasing the standard deviation of the normalized light curve from $0.053$ to $0.033$. 

TESS assigns a quality flag to all measurements. We removed all flagged data points except of `Impulsive outlier' and `Cosmic ray in collateral data' (bits 10 and 11) while extracting the light curve. The final light curve has $92839$ data points in two segments separated by the usual data downlink (so-called Low-Altitude Housekeeping Operations, LAHO)  gap. 
We applied a detrending to our cleaned light curve 
by removing a third order polynomial from both light curve segments individually. Then  the light curve was normalized.  


\section{Parameters from the spectral analysis of {\em XMM-Newton} data}\label{app:xrayfitres}

 We provide the elemental abundances for the quiescent corona of \adleo during the {\em XMM-Newton} observation derived from the RGS spectrum (Table~\ref{tab:rgs_abund}) and the evolution of the spectral parameters throughout the exponential flare phase obtained from the EPIC/pn spectra in time-slices of roughly equal photon-statistics (Table~\ref{tab:EPICfitres}) as explained in Sect.~\ref{subsubsect:analysis_xray_spectra}.

\begin{table}
    \centering
      \caption{Coronal abundances derived from the quiescent RGS spectrum with respect to the solar photospheric abundances from \cite{Asplund09.0} and first ionization potential.}
    \label{tab:rgs_abund}
    \begin{tabular}{lcr}\hline
    Element & $A_{\rm X}/A_{\rm X,\odot}$ & FIP \\ \hline
C  & $1.03^{+0.73}_{-0.13}$ & $11.3$ \\
N  & $1.36^{+0.38}_{-0.14}$ & $14.5$ \\
O  & $0.99^{+0.27}_{-0.04}$ & $13.6$ \\
Ne & $1.76^{+0.60}_{-0.09}$ & $21.6$ \\
Mg & $0.21^{+0.13}_{-0.13}$ & $7.6$ \\
Si & $0.66^{+0.30}_{-0.63}$ & $8.2$ \\
S  & $0.48^{+0.33}_{-0.41}$ & $10.4$ \\
Ar & $0.73^{+1.12}_{-0.66}$ & $15.8$ \\
Fe & $0.33^{+0.11}_{-0.02}$ & $7.9$ \\
\hline
    \end{tabular}
\end{table}

\begin{table*}
\caption{Best-fit parameters of the time-resolved X-ray spectral analysis of the EPIC/pn data throughout the exponential phase of the flare.}
\label{tab:EPICfitres}
\small
\begin{center}
\begin{tabular}{llr@{$\;\pm\;$}lr@{$\;\pm\;$}lr@{$\;\pm\;$}lr@{$\;\pm\;$}lr@{$\;\pm\;$}lr@{$\;\pm\;$}lr@{$\;\pm\;$}lr@{$\;\pm\;$}lr@{$\;\pm\;$}lr@{$\;\pm\;$}lr@{$\;\pm\;$}lr@{$\;\pm\;$}lr@{$\;\pm\;$}lr@{$\;\pm\;$}lr@{$\;\pm\;$}lr@{$\;\pm\;$}l}
\hline\hline
 int. & $t_{mid}^{a}$ &\multicolumn{2}{c}{ $\log T_{1}^{b}$ (K) } &\multicolumn{2}{c}{ $\log T_{2}^{b}$ (K) } &\multicolumn{2}{c}{ $\log EM_{1}^{b}$ (cm$^{-3}$) } &\multicolumn{2}{c}{ $\log EM_{2}^{b}$ (cm$^{-3}$) } &\multicolumn{2}{c}{ Fe/Fe$_{\odot}^{c}$ } & \multicolumn{2}{c}{ $\log T_{mean}^{d}$ (K) } &\multicolumn{2}{c}{ $\log EM_{tot}^{e}$ (cm$^{-3}$)} \\
\hline
  1 &  0.745 &   7.00 &   0.02 &   7.58 &   0.02 &  51.35 &   0.08 &  52.24 &   0.02 &   0.84 &   0.13 &   7.54 &   0.05 &  52.30 &   0.02 \\
  2 &  0.747 &   7.02 &   0.01 &   7.60 &   0.02 &  51.62 &   0.07 &  52.59 &   0.02 &   0.93 &   0.12 &   7.57 &   0.05 &  52.63 &   0.02 \\
  3 &  0.748 &   7.00 &   0.02 &   7.53 &   0.02 &  51.69 &   0.07 &  52.60 &   0.02 &   0.90 &   0.11 &   7.50 &   0.05 &  52.65 &   0.02 \\
  4 &  0.749 &   6.98 &   0.02 &   7.47 &   0.02 &  51.68 &   0.06 &  52.59 &   0.02 &   0.84 &   0.09 &   7.43 &   0.05 &  52.64 &   0.02 \\
  5 &  0.750 &   7.00 &   0.01 &   7.39 &   0.02 &  51.99 &   0.06 &  52.56 &   0.02 &   0.58 &   0.06 &   7.33 &   0.05 &  52.66 &   0.02 \\
  6 &  0.752 &   6.96 &   0.01 &   7.38 &   0.02 &  52.02 &   0.05 &  52.53 &   0.02 &   0.49 &   0.05 &   7.31 &   0.05 &  52.64 &   0.02 \\
  7 &  0.753 &   6.97 &   0.01 &   7.36 &   0.02 &  52.07 &   0.05 &  52.43 &   0.02 &   0.40 &   0.04 &   7.28 &   0.06 &  52.59 &   0.02 \\
  8 &  0.755 &   6.94 &   0.01 &   7.31 &   0.02 &  51.93 &   0.05 &  52.33 &   0.02 &   0.46 &   0.05 &   7.23 &   0.06 &  52.47 &   0.02 \\
  9 &  0.757 &   6.95 &   0.01 &   7.33 &   0.02 &  51.90 &   0.05 &  52.27 &   0.02 &   0.37 &   0.04 &   7.24 &   0.06 &  52.42 &   0.02 \\
 10 &  0.760 &   6.94 &   0.01 &   7.31 &   0.02 &  51.94 &   0.05 &  52.19 &   0.02 &   0.27 &   0.03 &   7.21 &   0.06 &  52.38 &   0.02 \\
 11 &  0.762 &   6.94 &   0.02 &   7.33 &   0.02 &  51.75 &   0.05 &  52.12 &   0.02 &   0.33 &   0.04 &   7.25 &   0.06 &  52.27 &   0.02 \\
 12 &  0.766 &   6.99 &   0.01 &   7.38 &   0.03 &  51.88 &   0.06 &  51.91 &   0.03 &   0.27 &   0.04 &   7.24 &   0.06 &  52.20 &   0.03 \\
 13 &  0.770 &   6.98 &   0.01 &   7.34 &   0.04 &  51.81 &   0.06 &  51.77 &   0.04 &   0.26 &   0.03 &   7.19 &   0.07 &  52.10 &   0.04 \\
 14 &  0.775 &   6.99 &   0.01 &   7.40 &   0.05 &  51.82 &   0.06 &  51.70 &   0.05 &   0.18 &   0.03 &   7.21 &   0.08 &  52.07 &   0.04 \\
 15 &  0.780 &   6.97 &   0.02 &   7.32 &   0.03 &  51.73 &   0.06 &  51.74 &   0.04 &   0.15 &   0.02 &   7.18 &   0.07 &  52.03 &   0.04 \\
 16 &  0.786 &   6.91 &   0.03 &   7.22 &   0.02 &  51.59 &   0.06 &  51.80 &   0.03 &   0.16 &   0.02 &   7.13 &   0.07 &  52.01 &   0.03 \\
 17 &  0.794 &   6.94 &   0.01 &   7.26 &   0.02 &  51.63 &   0.04 &  51.69 &   0.03 &   0.18 &   0.02 &   7.14 &   0.06 &  51.96 &   0.03 \\
\hline
\end{tabular}
\end{center}
$^a$~Time since BJD = 2459537.
$^b$~Results of the 2$T$ fit.
$^c$~Fe abundance.
$^d$~$EM$-weighted average temperature.
$^e$~Total $EM$.
\normalsize
\end{table*}

\section{Period search on the TESS light curve}\label{app:tess_rotation}

\begin{figure*}
\begin{center} 
\includegraphics[width=18.0cm]{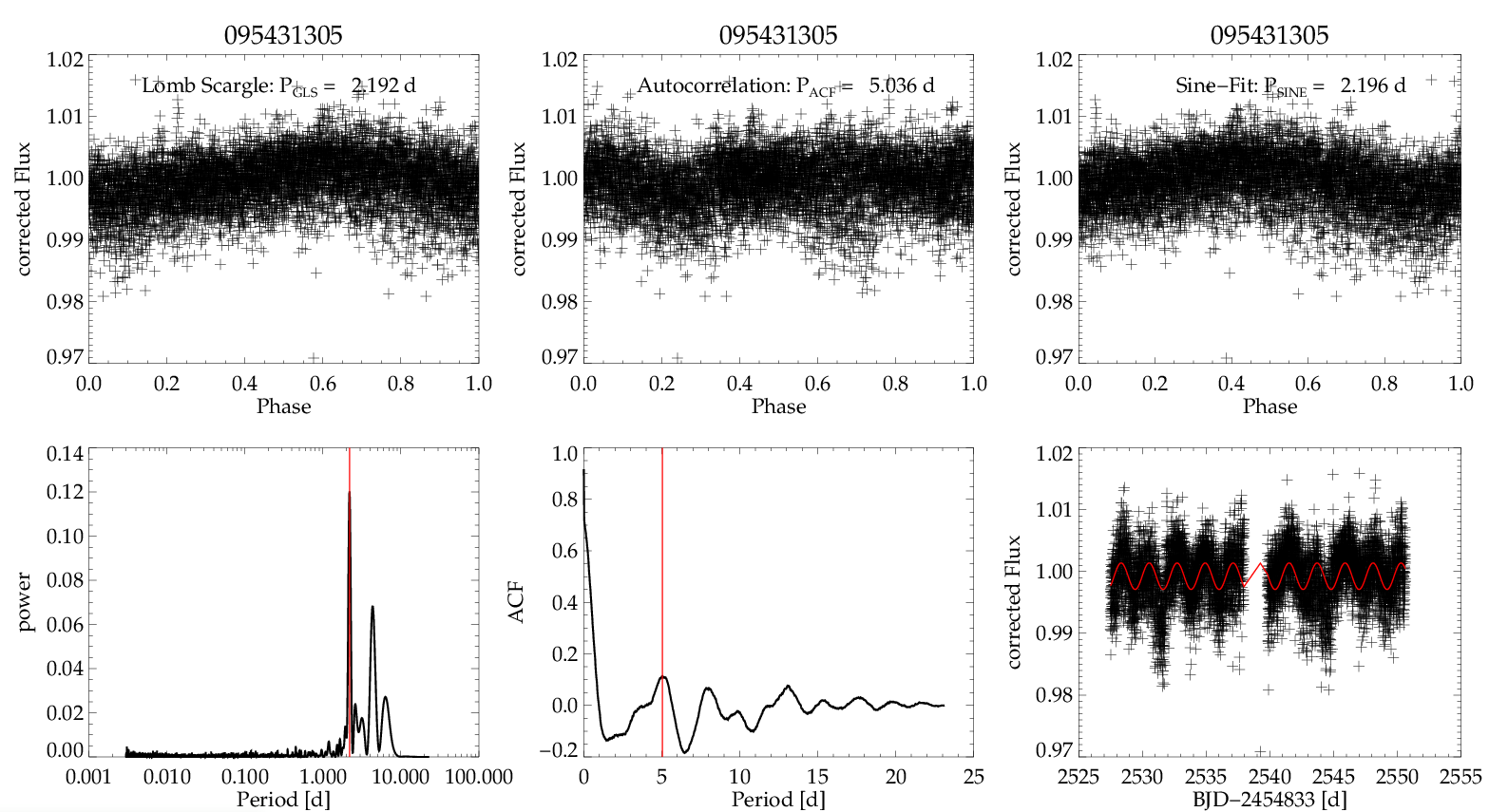}  
\caption{Results of the three period search methods (GLS, ACF and sine-fitting) for AD\,Leo
observed in TESS sector~45. The top panels show the light curve phase-folded with the periods
obtained with the different methods. The bottom panel shows the GLS periodogram, the ACF and
the original detrended light curve with the sine fit.} 
\label{fig:tess_rotation}
\end{center}
\end{figure*}

Following our previous work cited in Sect.~\ref{subsect:analysis_tess} 
we used three methods to search for the rotation period of AD\,Leo: We computed the generalized Lomb-Scargle periodogram \citep[\begin{small}GLS\end{small};][]{2009A&A...496..577Z}, we determined the autocorrelation function (ACF) and, finally, we fitted the light curve with a sine function. The \begin{small}GLS\end{small} implementation we used\footnote{Fortran Version v2.3.01, released 2011-09-13 by Mathias Zechmeister} can only process up to 10000 data points. Therefore, we had to bin the light curve by a factor of ten to a resolution of $200$\,s. The light curve was then phase-folded with the period found with each of the methods. The result of our period search is shown in Fig.~\ref{fig:tess_rotation}. Through visual inspection we selected the best-fitting period. For the {\em TESS} Sector\,45 light curve of AD\,Leo analysed in this work, the \begin{small}GLS\end{small} and the sine fitting resulted in periods consistent with each other and with values from the previous literature (see Sect.~\ref{subsubsect:analysis_tess_rotmod}), while the ACF did not show an unambiguous periodic pattern and, hence, failed to identify the rotation period 
(see Fig.~\ref{fig:tess_rotation}). We thus adopted the average of the \begin{small}GLS\end{small} and sine-fitting period, $P_{\rm rot} = 2.194 \pm 0.004$. The error was calculated with  
the formulas given by \citet{1985PASP...97..285G}.

\end{document}